\documentstyle[12pt,epsfig]{article}

\begin{document}

\renewcommand{\theequation}{\thesection .\arabic{equation}}
\renewcommand{\section}[1]{\setcounter{equation}{0}
 \addtocounter{section}{1}
 \vspace{5mm} \par \noindent {\large \bf \thesection . #1}
 \setcounter{subsection}{0} \par \vspace{2mm} } 
\newcommand{\sectionsub}[1]{\addtocounter{section}{1}
 \vspace{5mm} \par \noindent {\bf \thesection . #1}
 \setcounter{subsection}{0}\par}
\renewcommand{\subsection}[1]{\addtocounter{subsection}{1}
 \vspace{2.5mm}\par\noindent {\bf  \thesubsection . #1}\par
 \vspace{0.5mm}}
\renewcommand{\thebibliography}[1]{{\vspace{5mm}\par 
 \noindent{\large \bf References}\par \vspace{2mm}} 
 \list{$[$\arabic{enumi}$]$}{\settowidth\labelwidth{[#1]}
 \leftmargin \labelwidth \advance\leftmargin\labelsep
 \addtolength{\topsep}{-4em}\usecounter{enumi}}
 \def\newblock{\hskip .11em plus .33em minus .07em}
 \sloppy\clubpenalty4000\widowpenalty4000 \sfcode`\.=1000\relax 
 \setlength{\itemsep}{-0.4em}}
\newcommand{\acknowledgments}[1]{\vspace{5mm}\par 
 \noindent{\large \bf Acknowledgments}\par \vspace{2mm}}

\def\a{& \hspace{-7pt}}
\def\c{\hspace{-5pt}}
\def\bea{\begin{eqnarray}}
\def\eea{\end{eqnarray}}
\def\be{\begin{equation}}
\def\ee{\end{equation}}
\def\nn{\nonumber}
\def\alp{\alpha}
\def\bet{\beta}
\def\gam{\gamma}
\def\del{\delta}
\def\eps{\epsilon}
\def\sig{\sigma}
\def\lam{\lambda}
\def\Lam{\Lambda}
\def\m{\mu}
\def\n{\nu}
\def\r{\rho}
\def\s{\sigma}
\def\d{\delta}
\def\Z{{\bf Z}}
\def\e{\epsilon}
\def\st{\scriptstyle}
\def\mco{\multicolumn}
\def\epp{\epsilon^{\prime}}
\def\vep{\varepsilon}
\def\ra{\rightarrow}
\def\ab{\bar{\alpha}}
\def\dslash{\raisebox{1pt}{$\slash$} \hspace{-7pt} \partial}
\def\Dslash{\raisebox{1pt}{$\slash$} \hspace{-8pt} D}
\def\dslashs{\raisebox{0.75pt}{$\scriptstyle{\slash}$} \hspace{-4.5pt} 
 \partial}
\def\Dslashs{\raisebox{0.75pt}{$\scriptstyle{\slash}$} \hspace{-5.5pt} D}

\newcommand\ubar[1]{#1 \hspace{-5.5pt} 
 \raisebox{-2.5pt}{$\scriptscriptstyle{-}$}}
\newcommand\uubar[1]{#1 \hspace{-8pt} \raisebox{-2.75pt}{$\scriptstyle{-}$}}
\newcommand\uuubar[1]{#1 \hspace{-10pt} \raisebox{-3pt}{$\scriptstyle{-}$}
 \hspace{-4pt} \raisebox{-3pt}{$\scriptstyle{-}$ \vspace{10pt}}}
\newcommand\uuuubar[1]{#1 \hspace{-15pt} \raisebox{-5pt}{$\scriptstyle{-}$}
 \hspace{-4pt} \raisebox{-3pt}{$\scriptstyle{-}$ \vspace{10pt}}}

\newcommand{\sect}[1]{\setcounter{equation}{0} \section{#1}}
\newcommand{\eqn}[1]{(\ref{#1})}
\newcommand\rf[1]{(\ref{#1})}
\newcommand{\NPB}[3]{{Nucl.\ Phys.} {\bf B#1} (#2) #3}
\newcommand{\CMP}[3]{{Commun.\ Math.\ Phys.} {\bf #1} (#2) #3}
\newcommand{\PRD}[3]{{Phys.\ Rev.} {\bf D#1} (#2) #3}
\newcommand{\PLB}[3]{{Phys.\ Lett.} {\bf B#1} (#2) #3}
\newcommand{\JHEP}[3]{{JHEP} {\bf #1} (#2) #3}
\newcommand{\ft}[2]{{\textstyle\frac{#1}{#2}}\,}
\newcommand{\dt}{\partial_{\langle T\rangle}}
\newcommand{\dtbar}{\partial_{\langle\bar{T}\rangle}}
\newcommand{\al}{\alpha^{\prime}}
\newcommand{\mst}{M_{\scriptscriptstyle \!S}}
\newcommand{\mpl}{M_{\scriptscriptstyle \!P}}
\newcommand{\dv}{\int{\rm d}^4x\sqrt{g}}
\newcommand{\lv}{\left\langle}
\newcommand{\rv}{\right\rangle}
\newcommand{\ph}{\varphi}
\newcommand{\sbar}{\,\bar{\! S}}
\newcommand{\xbar}{\,\bar{\! X}}
\newcommand{\fbar}{\,\bar{\! F}}
\newcommand{\zbar}{\,\bar{\! Z}}
\newcommand{\tbar}{\bar{T}}
\newcommand{\ybar}{\bar{Y}}
\newcommand{\phb}{\bar{\varphi}}
\newcommand{\cm}{Commun.\ Math.\ Phys.~}
\newcommand{\pr}{Phys.\ Rev.\ D~}
\newcommand{\prl}{Phys.\ Rev.\ Lett.~}
\newcommand{\pl}{Phys.\ Lett.\ B~}
\newcommand{\ibar}{\bar{\imath}}
\newcommand{\jbar}{\bar{\jmath}}
\newcommand{\np}{Nucl.\ Phys.\ B~}
\newcommand{\eqalign}[1]{\hspace{-10pt}\begin{array}{ll} #1 
 \end{array}\hspace{-10pt}}
\newcommand{\dfrac}[2]{\displaystyle{\frac {#1}{#2}}}
\newcommand{\gsi}{\,\raisebox{-0.13cm}{$\stackrel{\textstyle>}
 {\textstyle\sim}$}\,}
\newcommand{\lsi}{\,\raisebox{-0.13cm}{$\stackrel{\textstyle<}
 {\textstyle\sim}$}\,}
\newcommand{\ds}[1]{\displaystyle{#1}}

\setcounter{page}{0}
\thispagestyle{empty}

\begin{center}
\hfill SISSA-94/2000/EP \\
\hfill LMU-TPW 00-23\\
\hfill{\tt hep-th/0010022}\\

\begin{center}

\vspace{1.7cm}

{\Large\bf A note on the torsion dependence \\

\vskip 5pt 

of D-brane RR couplings}

\end{center}

\vspace{1.4cm}

{\sc Claudio A. Scrucca$^{a,1}$ and Marco Serone$^{~b,2}$} \\

\vspace{1.2cm}

${}^a$
{\em Sektion Physik, Ludwig Maximilian Universit\"at M\"unchen}\\
{\em Theresienstra\ss e 37, 80333 Munich, Germany}\\
{\em E-mail: Claudio.Scrucca@physik.uni-muenchen.de}\\
\vspace{.3cm}

${}^b$
{\em Department of Mathematics, University of Amsterdam}\\
{\em Plantage Muidergracht 24, 1018 TV Amsterdam, The Netherlands} \\
{\em E-mail: serone@wins.uva.nl} \\
\vspace{.3cm}

\end{center}

\vspace{0.8cm}

\centerline{\bf Abstract}
\vspace{2 mm}
\begin{quote}\small

The dependence on the torsion $H=db$ of the Wess-Zumino couplings of 
D-branes that are trivially embedded in space-time is studied. 
We show that even in this simple set-up some torsion components
can be turned on, with a non-trivial effect on the RR couplings.
In the special cases in which either the tangent or the normal 
bundle are trivial, the torsion dependence amounts to substitute
the standard curvature with its generalization in the presence of torsion,
in the usual couplings involving the roof genus $\hat A$.

\end{quote}

\vfill
\begin{flushleft}
\rule{16.1cm}{0.2mm}\\[-3mm]
\end{flushleft}
$^1${\footnotesize Since 1$^{{\rm st}}$ October 2000:
CERN, 1211 Geneva 23, Switzerland. E-mail: Claudio.Scrucca@cern.ch}\\
$^2${\footnotesize Since 1$^{{\rm st}}$ September 2000:
ISAS-SISSA, Via Beirut 2-4, 34013 Trieste, Italy. E-mail: serone@sissa.it}

\newpage
\setcounter{equation}{0}

\section{Introduction}

By now, it is well known that both D(irichlet)-branes and 
O(rientifold)-planes must have Wess-Zumino (WZ) couplings to 
Ramond Ramond (RR) fields. These couplings are required for a 
consistent cancellation of anomalies, and their general form 
is well-known in quite arbitrary backgrounds \cite{pol}-\cite{ss1} 
(see \cite{mye} for effects peculiar to multiple branes). 
However, their dependence on the Neveu-Schwarz Neveu-Schwarz (NSNS)
$b$-field is not completely established, even for a single D-brane,
and its relevance has yet to be understood. At leading order in
derivatives ({\it i.e.} for constant $b$), the dependence of the 
WZ couplings on $b$ is completely fixed by the observation of 
\cite{wit1} that the gauge invariant field strength living on a 
D-brane is ${\cal F} = F - b/2\pi$, and has indeed been explicitly 
verified. On general grounds, one expects that the presence of
torsion ($H = db$) will modify these couplings, much on the same 
way as the presence of curvature does. For instance, it was pointed 
out in \cite{mye} that the restoration of a T-duality symmetric 
WZ coupling seems indeed to require a non-trivial dependence on the 
torsion, but no investigation has yet been done in this direction.

In this paper, we shall try to deduce the torsion-dependence of 
WZ couplings by factorizing one-loop CP-odd amplitudes, along the 
lines of \cite{mss,ss1,ss2}, although here we will consider only 
D-branes. More precisely, we will compute the inflow of anomaly 
arising from amplitudes with external curvature and torsion 
vertices. The one-loop amplitudes in question can be exponentiated,
reducing the problem to the evaluation of a twisted partition function
in the odd spin structure. In the case at hand, this will be the partition 
function of a supersymmetric $\sigma$-model in a curved and contorted 
background.

We consider D-branes that are trivially embedded in space-time. 
Requiring a trivial embedding, that is the usual Neumann (N) or 
Dirichlet (D) boundary conditions for the $\sigma$-model, puts 
severe constraints on the possible torsion terms that may appear. 
In particular the intrinsic torsion on both the tangent and normal 
bundle of D-branes must vanish, but some mixed components of $H$ 
are still allowed. They give rise to an antisymmetric part in the 
so-called second fundamental form defining the brane embedding in space-time 
(see the appendix). These are the torsion terms that will be studied in 
this paper. 

Unfortunately, we will not be able to derive a simple expression for the 
exact torsion dependence in the most general gravitational background,
which is nevertheless implicitly encoded in certain computable one-loop 
determinants\footnote{Strictly speaking, our results are derived for IIB 
D-branes but, as in \cite{ss1}, they apply to IIA D-branes as well.}. 
However, in the two particular cases of trivial tangent bundle and 
trivial normal bundle, we find that the torsion dependence simply amounts 
to the standard generalization of the curvature two-form in presence 
of torsion, as suggested in \cite{mye}\footnote{The simple replacement 
$R\rightarrow {\cal R}$ is known to occur for standard chiral fermions
in four dimensions \cite{ob,dm,pw} (see however \cite{cza} for a recent 
controversy), but a similar result in $D$ dimensions has been obtained
only for completely antisymmetric torsion \cite{mavro}.}.
Whether or not this extends to arbitrary backgrounds is not clear. 

Due to the limits of our analysis, the results of this paper should be
taken as a first effort to study and understand the torsion dependence of
D-brane couplings. It should also be pointed out that within this approach, 
non-anomalous torsion-dependent couplings cannot be detected. A more complete 
analysis is therefore needed to better understand these couplings. Also a 
more direct analysis along the lines of \cite{cr,stef} would be very 
interesting. Finally, D-branes in presence of torsion can be efficiently 
analysed within a different approach in the special case of group manifolds 
(see for instance \cite{fss} and references therein). This might be another 
helpful direction of investigation to better understand D-brane couplings.

\section{World-sheet theory}

As mentioned in the introduction, we analyse WZ couplings of D-branes by
factorizing anomalous one-loop amplitudes in the odd spin-structure.
Their torsion dependence can be studied by considering diagrams that
contain both gravitons and B-fields as external states.
In complete analogy to previous cases \cite{mss,ss1,ss2}, where the torsion
was set to zero, these amplitudes can be exponentiated.
In this way one extracts directly the polynomial of the inflow
of anomaly  that is given by the partition function (in the 
odd spin-structure) of the resulting $\sigma$-model.
By factorization, the WZ couplings responsible for this inflow of
anomaly can then be extracted without the need of implementing
the WZ descent procedure, that gives the actual gravitational or Lorentz
anomaly\footnote{See in particular section 2 of \cite{ss2} for further 
details.}.

The $\sigma$-model in question is the supersymmetric $\sigma$-model 
in presence of a generic gravitational and torsion background.
In superspace, the action is given by 
\be
S(\Phi) = \frac 14 \int\! d^2x \, d^2\theta \left[ g_{MN}(\Phi)
\bar D \Phi^M D \Phi^N - b_{MN}(\Phi) (\bar D \Phi^M \gamma^3 D \Phi^N)
\right] \;,
\label{SSM}
\ee
where $\Phi^M(x,\theta)=\phi^M (x) + \bar\theta \, \psi^M(x) +
1/2 \, \bar\theta \theta \, F^M(x)$ denote ten chiral superfields 
($M=0,...,9$),
$D_\alpha=\partial/\partial\, \theta^\alpha -i(\dslash \,\theta)_\alpha$
and $\bar D_\alpha$ is its complex conjugate. 
We take the following conventions for two-dimensional $\gamma$-matrices: 
in terms of Pauli matrices $\sigma_i$, $\gamma^0 = \sigma_2$,
$\gamma^1 = i \sigma_1$, $\gamma^3 = \sigma_3$. These satisfy the property
$\gamma^3 \gamma^\alpha = \epsilon^\alpha_{\;\;\beta} \gamma^\beta$, with
$\epsilon^{01}=+1$. Moreover, it is natural to introduce the following 
connections
\bea
\Gamma^M_{\;\;\,PQ} \a=\a \frac 12 \, g^{MN} \,
(g_{NP,Q} + g_{NQ,P} - g_{PQ,N}) \;, \label{Sconn} \\
H^M_{\;\;\,PQ} \a=\a \frac 12 \, g^{MN} \, 
(b_{NP,Q} + b_{PQ,N} + b_{QN,P}) \;, \label{Aconn}
\eea
and define the corresponding curvatures as
\bea
R^M_{\;\;\,NPQ} \a=\a \Gamma^M_{\;\;\,NQ,P} - \Gamma^M_{\;\;\,NP,Q} 
+ \Gamma^M_{\;\;\;RP} \, \Gamma^R_{\;\;\,NQ} 
- \Gamma^M_{\;\;\,RQ} \, \Gamma^R_{\;\;\,NP} \;, \label{R}\\
G^M_{\;\;\,NPQ} \a=\a H^M_{\;\;\,NQ;P} - H^M_{\;\;\,NP;Q} 
+ H^M_{\;\;\;RP} \, H^R_{\;\;\,NQ} 
- H^M_{\;\;\,RQ} \, H^R_{\;\;\,NP} \label{G} \;.
\eea
Commas and semicolons denote the usual derivatives and covariant derivatives,
in terms of the symmetric connection (\ref{Sconn}).

By expanding in $\theta,\bar \theta$ all the terms in (\ref{SSM}),
and eliminating the auxiliary fields $F^M$, one gets the following 
action in components \cite{cz}:
\bea
\a\a S = \frac 12 \int\! d^2x \left[ g_{MN} 
\partial_\alpha \phi^M \partial^\alpha \phi^N + \epsilon^{\alpha\beta}
b_{MN} \partial_\alpha \phi^M \partial_\beta \phi^N 
+ i g_{MN} \bar\psi^M \hat{\Dslash} \psi^N \right. \label{Sexp} \\
\a\a \hspace{67pt} \left. + \frac i2 \partial_\alpha (b_{MN} \bar\psi^M 
\gamma^\alpha\gamma^3 \psi^N ) + \frac 18 {\cal R}_{MNPQ} 
\bar\psi^M (1+\gamma^3) \psi^P \bar\psi^N (1+\gamma^3) \psi^Q \right] \nn
\eea
in terms of the generalized covariant derivative 
\be
\hat{\Dslash} \psi^M = \gamma^\alpha [\partial_\alpha \psi^M 
+ (\Gamma_{\;\;\,PQ}^M - \gamma^3 H_{\;\;\,PQ}^M) \, 
\partial_\alpha \phi^P \psi^Q ] \;,
\ee
and the generalized Riemann tensor 
\be
{\cal R}_{MNPQ}= R_{MNPQ} + G_{MNPQ} \label{calR} \;.
\ee
constructed from $\Gamma^M_{\;\;\,PQ} + H^M_{\;\;\,PQ}$ \cite{ss}.
The action (\ref{Sexp}) is invariant under the following supersymmetry 
transformations:
\bea
\delta_\epsilon \phi^M \a=\a \bar \epsilon \psi^M \;, \nn \\
\delta_\epsilon \psi^M \a=\a - i \dslash \phi^M \epsilon
+ \frac 12 (\Gamma^M_{\;\;\,AB} \bar \psi^A \psi^B 
+ H^M_{\;\;\,AB} \bar \psi^A \gamma^3 \psi^B) \epsilon \;. \label{susy}
\eea

From a $\sigma$-model point of view, D-branes are represented as 
world-sheet boundaries with suitable Neumann (N) or Dirichlet (D) 
boundary conditions (b.c.). D-branes in flat spaces or trivially embedded 
in curved space satisfy the usual b.c.:
\bea
\a\a \partial_\sigma \phi^\mu(0,\tau) =0 \;,\hspace{1.2cm}
\partial_\tau \phi^i(0,\tau) = 0 \;, \nn \\
\a\a \psi_1^\mu (0,\tau) = \psi_2^\mu (0,\tau) \;,\;\;\; 
\psi_1^i(0,\tau) = -\psi_2^i(0,\tau) \;,
\label{bc}
\eea
where $\psi_1$ and $\psi_2$ are the two components of the Majorana
spinor $\psi$.
Here and throughout the paper we use greek indices $\mu,\nu,...=0,...,p$  
and latin indices $i,j,... = p+1,...,9$ to denote respectively N and D
directions of a Dp-brane.
We now implement the b.c. (\ref{bc}) in the action (\ref{Sexp}) and require,
as usual, that all the boundary terms in the variation of (\ref{Sexp}) 
vanish. It is a simple although laborious exercise to verify that
all boundary terms vanish if the following constraints on the background 
are satisfied:
\bea
g_{\mu i}\mid_M \a = \a \partial_i g_{\mu\nu}\mid_M = 0 \;, \nn \\
b_{\mu\nu}\mid_M \a = \a b_{ij}\mid_M = \partial_i \,b_{\mu j}\mid_M =0 \;,
\label{res2}
\eea
where $\mid_M$ is to remind that these conditions must hold only on 
the boundary of the world-sheet, that is the D-brane world-volume $M$.
In terms of the field strength $H$, 
(\ref{res2}) imply that $H_{\mu\nu\rho}=H_{\mu ij}=H_{ijk}=0$.
In other words, the only possible torsion components compatible with 
the usual b.c. (\ref{bc}) are those with two N and one D indices, 
$H_{\mu\nu i}$\footnote{There is also an alternative and probably 
faster way to see how the conditions (\ref{res2}) arise.
By taking the usual vertex operator for B one easily verify, 
using the b.c. conditions (\ref{bc}), that on $M$ only the 
$H_{\mu\nu i}$ components survive.}.

Generically, a world-sheet boundary also breaks both of the two world-sheet 
supersymmetries. This is indeed the case for the conditions (\ref{bc}), in 
a generic background. Interestingly, the same constraints (\ref{res2}) are 
required to leave a combination of the left and right supersymmetries 
unbroken. In fact, if (\ref{res2}) hold, the combination 
$\delta = \delta_1 - \delta_2$ of the two original supersymmetry 
variations $\delta_{1,2}$ is preserved.

Notice finally that (\ref{res2}) are not the only possible solution to the 
boundary conditions (\ref{bc}). One could also consider more complicated cases 
where different terms in (\ref{res2}) are non-vanishing and compensate each 
other to give a total vanishing boundary term in the variation of the action 
(\ref{Sexp}). We will not consider such cases.

\section{Reduction to 0+1 dimension and quantization.}

When the world-sheet theory is supersymmetric, the evaluation of the
partition function in the odd spin-structure is greatly simplified.
Indeed, it is a topological quantity, the Witten index \cite{wit},
receiving contributions only from zero-energy states. 
These correspond to field configurations
which are constant in the space-like direction of the world-sheet, and 
one can therefore use a 0+1 dimensional effective theory for the computation
of the index. 

It is convenient to introduce fermions with flat indices both on the 
tangent and the normal bundles. Defining then new fermions as 
$\psi^{\ubar{\mu}} = e^{\ubar{\mu}}_\mu \psi^\mu$ and 
$\psi^{\ubar{i}} = e^{\ubar{i}}_i \psi^i$, where
$\psi^\mu = \psi_1^\mu/\sqrt{2} = \psi_2^\mu/\sqrt{2}$ and 
$\psi^i = \psi_1^i/\sqrt{2} = - \psi_2^i/\sqrt{2}$,
the $1+0$ dimensional reduction of the action (\ref{Sexp}),
with the restrictions (\ref{res2}), yields:
\bea
\a\a L = \frac 12 g_{\mu\nu} 
\dot \phi^\mu \dot \phi^\nu 
+ \frac i2 \psi^{\ubar{\mu}} \Big(\dot \psi_{\ubar{\mu}}
+ \omega^{(0)}_{\rho\,\ubar{\mu}\ubar{\nu}} 
\dot \phi^\rho \psi^{\ubar{\nu}} \Big) 
+ \frac i2 \psi^{\ubar{i}} \Big(\dot \psi_{\ubar{i}}
+ \omega^{(0)}_{\rho\,\ubar{i}\ubar{j}} 
\dot \phi^\rho \psi^{\ubar{j}} \Big)+ \frac 14 
R_{\ubar{\mu}\ubar{\nu}\ubar{i}\ubar{j}} 
\psi^{\ubar{\mu}} \psi^{\ubar{\nu}} \psi^{\ubar{i}} \psi^{\ubar{j}}  \nn \\
\a\a \hspace{25pt} - \,i H_{\rho\ubar{\mu}\ubar{i}} \dot \phi^\rho 
\psi^{\ubar{\mu}}\psi^{\ubar{i}}
-\frac 16 H_{\ubar\mu \ubar\nu\ubar\rho ;\,\ubar i} 
\psi^{\ubar{\mu}} \psi^{\ubar{\nu}} \psi^{\ubar{\rho}} \psi^{\ubar i} 
+ \frac 18 H_{\ubar k\ubar{\mu}\ubar{\nu}} 
H^{\ubar k}_{\;\;\,\ubar{\rho}\ubar{\sigma}}
\psi^{\ubar{\mu}} \psi^{\ubar{\nu}} \psi^{\ubar{\rho}} 
\psi^{\ubar{\sigma}} \;, \label{Aflat}
\eea
where $\omega^{(0)}$ is the torsion-free connection one-form.
The supersymmetry transformations leaving it invariant are
($\epsilon = \epsilon_1/\sqrt{2} = - \epsilon_2/\sqrt{2}$):
\bea
\delta_\epsilon \phi^\mu \a=\a 
i \,e^\mu_{\ubar{\mu}} \,\psi^{\ubar{\mu}} \,\epsilon \;, \nn \\
\delta_\epsilon \psi^{\ubar{\mu}} \a=\a  
e^{\ubar{\mu}}_\mu \, \dot \phi^\mu \, \epsilon  
+ i \, e_{\ubar{\nu}}^\nu \, 
\omega^{(0)\ubar\mu}_{\,\nu\;\;\;\;\ubar{\rho}} 
\psi^{\ubar{\nu}} \psi^{\ubar{\rho}} \, \epsilon \;, 
\raisebox{15pt}{} \nn \\
\delta_\epsilon \psi^{\ubar{i}} \a=\a 
i \, e^\mu_{\ubar{\mu}}\, \omega^{(0)\ubar i}_{\mu\;\;\;\;\ubar{j}}
\,\psi^{\ubar{\mu}} \psi^{\ubar{j}} \, \epsilon 
+ \frac i2 H^{\ubar i}_{\;\,\ubar{\mu}\ubar{\nu}} 
\psi^{\ubar{\mu}} \psi^{\ubar{\nu}} \epsilon \;,
\label{susyflat}
\eea
and the supercharge is
\be
Q = e_{\mu\ubar{\nu}} \dot \phi^\mu \psi^{\ubar{\nu}} 
- \frac i2 H_{\ubar{\mu}\ubar{\nu} \ubar{i}} 
\psi^{\ubar{\mu}} \psi^{\ubar{\nu}} \psi^{\ubar{i}} \label{Q}\;.
\ee

Before starting to evaluate the partition function associated to the action
(\ref{Aflat}), it is very useful to quantize the theory to have a more
precise understanding of which kind of anomalies we are studying.
Indeed, it is well-known that the ill-defined traces encoding anomalies
in Fujikawa's approach can be regulated and evaluated as the high temperature 
limit of the partition functions of suitable supersymmetric theories, 
corresponding to indices of certain operators.
An investigation in this direction is also further motivated by the 
observation of \cite{mavro} that the Atiyah-Singer index theorem in 
presence of torsion is associated to a supersymmetric quantum mechanical 
model that is {\it not} the reduction to 0+1 dimension of (\ref{Sexp}) 
with the constraints (\ref{bc})\footnote{
When the torsion vanishes, these are the standard constraints to
compute pure gravitational anomalies \cite{agw} (see \cite{ss1} for
the relevance of Dirichlet boundary conditions for non-trivial
normal bundles) whose form is also given by the Atiyah-Singer
index theorem.}; the right model is rather the reduction to 0+1 
dimension of an heterotic $\sigma$-model.
This is in agreement with our result that for the purely Neumann case, 
no torsion is consistent with the conditions (\ref{bc}).

The conjugate momenta for $\phi^\mu$, $\psi^{\ubar{\mu}}$
and $\psi^{\ubar{i}}$ are given by
\be
\pi_\mu = g_{\mu\nu} \dot \phi^\nu 
+ \frac i2 \omega^{(0)}_{\mu\,\ubar{\alpha}\ubar{\beta}} 
\psi^{\ubar{\alpha}}\psi^{\ubar{\beta}} 
+ \frac i2 \omega^{(0)}_{\mu\,\ubar{i}\ubar{j}} 
\psi^{\ubar{i}} \psi^{\ubar{j}} - i H_{\mu\ubar{\nu}\ubar{i}} 
\psi^{\ubar{\nu}} \psi^{\ubar{i}} \;,\;\; 
\tau_{\ubar{\mu}} = \frac i2 \psi_{\ubar{\mu}} \;,\;\;
\tau_{\ubar{i}} = \frac i2 \psi_{\ubar{i}} \;,
\ee
and the Hamiltonian is 
$$
H = \frac 12 g_{\mu\nu} \dot \phi^\mu \dot \phi^\nu 
- \frac 14 R_{\ubar{\mu}\ubar{\nu}\ubar{i}\ubar{j}} 
\psi^{\ubar{\mu}} \psi^{\ubar{\nu}} \psi^{\ubar{i}} \psi^{\ubar{j}} 
+ \frac 16 H_{\ubar\mu \ubar\nu\ubar\rho ;\,\ubar i} 
\psi^{\ubar{\mu}} \psi^{\ubar{\nu}} \psi^{\ubar{\rho}} \psi^{\ubar i}
- \frac 18 H_{k\ubar{\mu}\ubar{\nu}} H^k_{\;\;\,\ubar{\rho}\ubar{\sigma}}
\psi^{\ubar{\mu}} \psi^{\ubar{\nu}} \psi^{\ubar{\rho}} 
\psi^{\ubar{\sigma}} \;.
$$
The Hamiltonian formulation and the quantization of supersymmetric 
quantum mechanical models like the one studied here presents some 
well-known subtleties. Indeed, due to the constraints in the fermionic 
sector of phase-space, it is necessary to use the Dirac procedure, and 
replace the standard Poisson brackets with Dirac brackets which are 
compatible with these constraints. Quantization can then proceed in the 
usual way, replacing Dirac brackets with commutators or anticommutators
(see \cite{dmph,bcz,braden}). We will assume here that the net result of 
this lengthy procedure is that one can use as canonical variables 
$\phi^\mu$, $\pi_\mu$, $\psi^{\ubar{\mu}}$ and $\psi^{\ubar{i}}$, with 
the following non-vanishing commutations relations:
\be
[\phi^\mu, \pi_\nu ] = i \delta^\mu_\nu \;,\;\;
\{\psi^{\ubar{\mu}}, \psi^{\ubar{\nu}} \} = \eta^{\ubar{\mu}\ubar{\nu}} 
\;,\;\; 
\{\psi^{\ubar{i}}, \psi^{\ubar{j}} \} = \delta^{\ubar{i}\ubar{j}} \;.
\label{comm}
\ee
In terms of these variables, the supercharge becomes:
\be
Q = - i e^\mu_{\ubar{\mu}} \psi^{\ubar{\mu}} \Big(i \pi_\mu 
+ \frac 12 \omega^{(0)}_{\mu\,\ubar{\rho}\ubar{\sigma}} 
\psi^{\ubar{\rho}} \psi^{\ubar{\sigma}} 
+ \frac 12 \omega^{(0)}_{\mu\,\ubar{i}\ubar{j}} 
\psi^{\ubar{i}} \psi^{\ubar{j}}
- \frac 12 H_{\mu\ubar{\rho} \ubar{i}} 
\psi^{\ubar{\rho}} \psi^{\ubar{i}} \Big) \;,
\label{Qfs}
\ee
and it is straightforward to check that $Q$ does indeed generate the correct
supersymmetry transformations (\ref{susyflat}): 
$\delta_\epsilon = [\epsilon Q,\;\;]$. Also, the supersymmetry algebra 
guarantees that $H = 1/2\{Q,Q\}$.

According to (\ref{comm}), the canonical operators $\phi^\mu$, $\pi_\mu$, 
$\psi^{\ubar{\mu}}$ and $\psi^{\ubar{i}}$ can be realized on the target 
space as
\be
\phi^\mu \rightarrow x^\mu \;,\;\; \pi_\mu \rightarrow -i \partial_\mu 
\;,\;\; \psi^{\ubar{\mu}} \rightarrow \gamma^{\ubar{\mu}}/\sqrt{2} 
\;,\;\; \psi^{\ubar{i}} \rightarrow \gamma^{\ubar{i}}/\sqrt{2} \;,
\label{tsp}
\ee
where $\gamma^{\ubar\mu},\gamma^{\ubar i}$ are the space-time 
$\gamma$-matrices in the directions which are respectively parallel and 
transverse to the brane. 
The supercharge (\ref{Qfs}) is then finally given by $Q = -i\Dslash/\sqrt{2}$, 
where $\Dslash$ is the following world-volume Dirac operator:
\be
\Dslash = e^\mu_{\ubar{\mu}} \gamma^{\ubar{\mu}} \Big(\partial_\mu 
+ \frac 14 \omega^{(0)}_{\mu\,\ubar{\rho}\ubar{\sigma}} 
\gamma^{\ubar{\rho}\ubar{\sigma}} 
+ \frac 14 \omega^{(0)}_{\mu\,\ubar{i}\ubar{j}} \gamma^{\ubar{i}\ubar{j}}
- \frac 14 H_{\mu\ubar{\rho}\ubar{i}} \gamma^{\ubar{\rho}}
\gamma^{\ubar{i}} \Big) \;.
\label{Dirac}
\ee
The operator (\ref{Dirac}) contains as expected a mixed torsion
connection, beside the usual tangent and normal bundle spin connections.
This shows that the index computed here encodes the anomaly 
of a chiral spinor in a curved and contorted background.

\section{Inflow of anomaly and WZ couplings}

We now turn to the computation of the inflow of anomaly.
This section follows closely the analysis reported in \cite{ss1},
with some modifications due to the presence of torsion.
According to the previous considerations, the inflow of anomaly is given by
the high temperature limit of the partition function 
\be
Z = \mbox{Tr} \,[\Gamma^{D+1} e^{- t (i \Dslashs)^2}] 
\;,
\label{anospinor}
\ee
where $\Gamma^{D+1}$ is the chiral matrix in $D$ dimensions
and $\Dslash$ is the Dirac operator (\ref{Dirac}).

The functional integral representation for (\ref{anospinor}) is
\be
Z = \int_P \! {\cal D} \phi^\mu(\tau)  \int_P \! {\cal D} \psi^{\ubar\mu} 
(\tau) \int_P {\cal D} \! \psi^{\ubar i}(\tau) 
\exp \left\{- \int_0^t \! d\tau \, L \left(\phi^\mu(\tau),\psi^{\ubar \mu} 
(\tau),\psi^{\ubar i}(\tau) \right) \right\} \;,
\label{pathspinor}
\ee
with $L$ as in (\ref{Aflat}).
All the fields are periodic $(P)$ in the odd spin-structure. 
In order to evaluate this path-integral in the high-temperature limit 
$t \rightarrow 0$, it is convenient to expand the fields in normal
coordinates around constant paths $\phi^\mu = \phi_0^\mu + \xi^{\mu}$, 
$\psi^{\ubar \mu} = \psi_0^{\ubar \mu} + \chi^{\ubar \mu}$ and 
$\psi^{\ubar i} = \psi_0^{\ubar i} + \chi^{\ubar i}$.

By doing the expansion described above, one should pay attention
to dangerous terms involving four Neumann fermionic zero modes 
and leading to divergences. A term of this type, involving the standard
Riemann tensor, has already been dropped because it vanishes thanks to the
Bianchi identity, but the last term in (\ref{Aflat}) remains, essentially
as a consequence of the fact that the torsion part of the curvature (\ref{G}) 
does not satisfy the same Bianchi identity as the geometric part (\ref{R}).
Here, we shall assume a very safe approach and restrict to the particular
case in which
\be
H_{\ubar i [\ubar\mu \ubar\nu} 
H^{\ubar i}_{\;\; \ubar\rho \ubar\sigma ]}= 0 \;.
\label{BianchiH}
\ee
In this case, for the same arguments explained in subsection 
2.1.1 of \cite{ss1}, it is sufficient to keep only interaction terms 
up to quadratic order in the fluctuations and, among these, only those 
involving fermionic zero modes $\psi_0^{\ubar\mu}$ in the Neumann
directions. One gets then the following effective Lagrangian:
\be
L^{eff}\! = 
\frac 12 \Big[\dot \xi_{\ubar \mu} \dot \xi^{\ubar \mu} 
\!+\! i \chi_{\ubar \mu} \dot \chi^{\ubar \mu} 
\!+\! i \chi_{\ubar i} \dot \chi^{\ubar i} 
\!+\! i R_{\ubar \mu \ubar \nu} \xi^{\ubar \mu} \dot \xi^{\ubar \nu}
\!+\! R_{\ubar i \ubar j} \chi^{\ubar i} \chi^{\ubar j} 
\!+\! 2i H_{\ubar \mu \ubar i} \dot \xi^{\ubar\mu} \chi^{\ubar i}
\!+\! R_{\ubar i \ubar j} \psi_0^{\ubar i} \psi_0^{\ubar j} \Big]
\label{spinoreff}
\ee
with\footnote{In the following, we shall distinguish through a prime
the normal bundle curvature from the tangent bundle curvature.}
\be
R_{\ubar\mu\ubar\nu} = \frac 12 R_{\ubar\mu\ubar\nu\ubar\rho\ubar\sigma} 
(\phi_0) \psi_0^{\ubar \rho} \psi_0^{\ubar \sigma} \;, \;\;\;
R_{\ubar i\ubar j} = \frac 12 R_{\ubar i \ubar j \ubar \rho \ubar \sigma} 
(\phi_0) \psi_0^{\ubar \rho} \psi_0^{\ubar \sigma} \;, \;\;\;
H_{\ubar \mu \ubar i} = H_{\ubar \mu \ubar i \ubar \rho} (\phi_0) 
\psi_0^{\ubar \rho} \;.
\label{tncurv}
\ee

Since the effective Lagrangian (\ref{spinoreff}) is quadratic, it is 
in principle straightforward at this point to compute the partition 
function (\ref{pathspinor}). As expected, the result is independent
of $t$, as can be seen by rescaling the Neumann fermionic zero modes.
As in the torsionless case, the last term in (\ref{spinoreff}) produces 
the Euler class of the normal bundle $e(R^\prime)$. The remaining pieces
of the Lagrangian give a bunch of determinants, which for the time being
we denote by $Y^2(R,R^\prime,H)$. The result is therefore:
\be
Z = \int_M Y^2(R,R^\prime,H) \, e(R^\prime) \;.
\label{Aspi}
\ee
As noted in \cite{cy}, the Euler class term in (\ref{Aspi})
is due to a topological property of currents and it appears
only in the inflow of anomaly, and must not be taken into account 
when factorizing (\ref{Aspi}) to extract the WZ couplings for 
D-branes\footnote{There is a subtlety here for the D3-brane \cite{cy}. In 
this case the anomaly ${\cal A}=2\pi i Z^{(1)}=2\pi i\int_M e(R^\prime)^{(1)}$ 
does not vanish, whereas the inflow of anomaly does, since the descent 
procedure for the inflow has not to be taken to the Euler class term.
We do not have a resolution to this issue, that seems to be due to 
additional subtleties in the definition of the WZ coupling of the 
self-dual D3-brane \cite{cy}.}.
Once $Y$ is known, the D-brane WZ couplings we are looking for are given by
\be
S_{D_p} = \frac {\mu_{p}}{2} \int_M \left.\,C\wedge\, Y
\right|_{(p+1)-form} \;,
\label{WZD} 
\ee
using the same conventions of \cite{ss1} to normalize the $D_p$ brane
charge $\mu_p$ and denoting with $C$ the formal sum of all the RR forms.

In spite of the fact that the effective Lagrangian (\ref{spinoreff}) is 
quadratic, the evaluation of the corresponding determinants is difficult, 
due to the presence of a mixing between bosons and fermions. The general 
result for $Y$ does not seem to lead to any simple combination of 
characteristic classes. This is quite disappointing but actually 
perfectly sensible, and probably just reflects the well-known difficulties 
in incorporating unambiguously the effects of torsion in characteristic 
classes. Rather than insisting on the exact result, we will limit ourselves 
to the two particular cases in which either the tangent or the normal bundle 
is trivial, which lead to simple results.

In the case of generic tangent bundle but trivial normal bundle, 
$R_{\ubar i\ubar j} = 0$, it is convenient to redefine the 
fermionic fluctuation as $\chi_{\ubar i} \rightarrow \chi_{\ubar i} 
+ i H_{\ubar i \ubar \mu} \xi^{\ubar \mu}$. By doing so, the torsion 
term in (\ref{spinoreff}) gets reabsorbed and an effective bosonic 
interaction is generated:
\be
L^{eff} = \frac 12 \left[\dot \xi_{\ubar \mu} \dot \xi^{\ubar \mu}
+ i \chi_{\ubar \mu} \dot \chi^{\ubar \mu} 
+ i \chi_{\ubar i} \dot \chi^{\ubar i} + i\left(R_{\ubar \mu \ubar \nu} 
- H_{\ubar i \ubar \mu} H^{\ubar i}_{\;\,\ubar \nu} \right)
\xi^{\ubar \mu} \dot \xi^{\ubar \nu} \right] \;.
\label{spinoreff2}
\ee
The evaluation of the determinants is then straightforward, and one finds:
\be
Y = \sqrt{\widehat A ({\cal R})} \;,
\ee
in terms of the generalized curvature of the tangent bundle, 
eq.(\ref{Gauss1}).

In the case of trivial tangent bundle, $R_{\ubar\mu\ubar\nu} = 0$,
but generic normal bundle, it is instead more convenient to redefine 
the bosonic fluctuations $\xi^{\ubar\mu}$ in such a way that 
$\dot \xi_{\ubar\mu} \rightarrow \dot \xi_{\ubar\mu} 
+ i H_{\ubar\mu\ubar i} \chi^{\ubar i}$. By doing this, the torsion 
term in (\ref{spinoreff}) gets again reabsorbed and this time an 
effective fermionic interaction is generated:
\be
L^{eff} = \frac 12 \left[\dot \xi_{\ubar \mu} \dot \xi^{\ubar \mu} 
+ i \chi_{\ubar \mu} \dot \chi^{\ubar \mu} 
+ i \chi_{\ubar i} \dot \chi^{\ubar i} + \left(R_{\ubar i \ubar j} 
- H_{\ubar\mu\ubar i} H^{\ubar\mu}_{\;\,\ubar j} \right)
\, \chi^{\ubar i} \chi^{\ubar j}
+ R_{\ubar i \ubar j} \psi_0^{\ubar i} \psi_0^{\ubar j} \right]
\label{spinoreff1} \;.
\ee
One finds then:
\be
Y = \sqrt \frac 1{\widehat A ({\cal R}^\prime)} \;,
\ee
in terms of the generalized curvature of the normal bundle, 
eq.(\ref{Ricci1}).

\section{Conclusions}

In this note, we have shown that D-brane anomalous couplings 
do have a non-trivial dependence on torsion. The precise 
form of this dependence is implicitly encoded in certain 
one-loop determinants, but does apparently not admit any simple 
expression in terms of standard characteristic classes. Specialising
to the two cases of trivial tangent and normal bundles, we were however
able to prove that the torsion dependence amounts simply to the 
replacement $R\rightarrow {\cal R}$ and $R^\prime\rightarrow {\cal R}^\prime$ 
in the usual torsion-free RR couplings. One might then wonder whether,
as guessed in \cite{mye}, the result for generic curvature and 
torsion is miraculously:
\be
Y = \sqrt{\frac {\widehat A ({\cal R})}{\widehat A ({\cal R}^\prime)}}
\;.
\ee
Unfortunately, we are not able to answer unambiguously to this question
since in this general case we did not find a sensible way to compute the 
determinants arising from the Lagrangian (\ref{spinoreff}). Furthermore,
we will not adress here the issue of whether the couplings we find restore 
T-duality, as proposed in \cite{mye}, because the restricted region of 
validity of our results does not allow to make any precise statement in 
this direction.

We also find a bit problematic to relax the condition (\ref{BianchiH})
on the torsion. We did not analyze in full detail the consequence 
of the presence of such a term in (\ref{Aflat}), but it is not excluded
that the potential divergence we found is linked to similar 
divergent terms appearing in the literature for the computation of 
the four-dimensional chiral anomaly in presence of torsion\footnote{It
might be that this divergent term is irrelevant upon integration over 
the world-volume.}.

An important observation is now in order. As shown in \cite{hbpt},
torsion does not give rise to new true anomalies, as long as it appears 
just through generalized curvatures in the usual characteristic classes.
Indeed, the addition of a covariant term to the spin-connection 
gives rise to a modification to the gravitational anomaly that may be 
reabsorbed by adding to the action a local counter-term.
This suggests that there could be some ambiguity in deriving results
in this context using anomaly arguments. However, the quantity we 
compute can be interpreted as the inflow of anomaly arising from the 
RR interaction of D-branes with couplings (\ref{WZD}). As such, we believe 
that our results are not affected by this ambiguity. 
We stress this important point because, differently from the 
torsion-free case, the supersymmetric quantum mechanical models 
relevant to compute the inflow of anomaly on D-branes and generic anomalies 
with torsion are different, as already mentioned in section 3. 

\vspace{5mm}
\par \noindent {\large \bf Acknowledgments}
\vspace{3mm}

We would like to thank L. Bonora and R. Dijkgraaf for very interesting 
discussions. This work has been partially supported by the EC under 
TMR contract ERBFMRX-CT96-0045, the Fundamenteel Onderzoek der Materie 
(FOM), the International School for Advanced Studies (ISAS-SISSA) and 
the European Organization for Nuclear Research (CERN).

\vspace{1cm}
\renewcommand{\theequation}{A.\arabic{equation}}
\setcounter{equation}{0}
\par \noindent{\large \bf A. Geometry of sub-manifolds with torsion}
\vspace{3mm}

In this appendix we give a brief review of the geometry of
sub-manifolds, along the lines of standard books \cite{oku-kn}. 
A good reference is also \cite{eis}.
A similar analysis for the torsion-free case,
and applied again to D-brane physics, can also be found in 
appendix A of \cite{bbg}.

Let $X$ be the ten-dimensional space-time endowed with a generic connection
and let $M$ be a $p+1$-dimensional sub-manifold of $X$, corresponding to the
embedding of the D-brane world-volume into space-time.
The embedding is defined by the equations 
$\phi^M=\phi^M(\sigma^\mu )$ ($\mu = 0,...,p$).
Cartan's structure equations on $X$ read then
\bea
\a\a T^{\underline M} = d\theta^{\underline M} + 
\omega^{\underline M}_{\;\;\;\underline N} 
\wedge \theta^{\underline N} \;, \label{CT} \\
\a\a {\cal R}^{\underline M}_{\;\;\;\underline N} = 
d\omega^{\underline M}_{\;\;\;\underline N}
+ \omega^{\underline M}_{\;\;\;\ubar P} \wedge 
\omega^{\ubar P}_{\;\;\;\underline N} \;, 
\label{CR}
\eea
where underlined indices represent flat indices and we denoted
the two-form curvature with ${\cal R}$ to distinguish it from the
geometric curvature $R$, constructed in terms of the torsion-free
connection form $\omega^{(0)}$. 

It is always possible to choose an orthonormal frame, the so-called 
``adapted frame'' \cite{oku-kn}, in which 
$\theta^{\ubar i}\mid_M=0$ $(\ubar i = p+1, ..., 9)$.
In such a frame, (\ref{CT}) with 
$\underline{\mbox{\footnotesize M}}= \ubar i$ yields 
$\omega^{\ubar i}_{\;\;\;\ubar \nu} \wedge \theta^{\ubar \nu} = T^{\ubar i}$.
By writing $\omega^{\ubar i}_{\;\;\;\ubar \nu} = 
\Omega^{\ubar i}_{\;\; \;\ubar\nu \ubar\mu} \theta^{\ubar\mu}$
and using the explicit form in components for the torsion form,
$T^{\ubar i}=  H^{\,\ubar i}_{\;\; \;\ubar\mu \ubar\nu} 
\theta^{\ubar\mu} \wedge \theta^{\ubar\nu}$, this can be rewritten as
\be
\left( \Omega^{\ubar i}_{\;\;\; \ubar\nu \ubar\mu}
+  H^{\,\ubar i}_{\;\;\; \ubar\mu \ubar\nu} \right)
\theta^{\ubar\mu} \wedge \theta^{\ubar\nu} = 0 \;.
\label{CTM}
\ee
The most generic solution for (\ref{CTM}) is
$\Omega^{\ubar i}_{\;\;\; \ubar\mu \ubar\nu} =
\Omega^{\ubar i}_{\;\;\; (\ubar\mu \ubar\nu)} +
H^{\,\ubar i}_{\;\;\; \ubar\mu \ubar\nu}$.
The tensor $\Omega^{\underline i}_{\;\; \ubar\mu \ubar\nu}$
is called the second fundamental form and plays an important role
in relating tensors on a manifold with those defined in its sub-manifolds.

From (\ref{CR}) with $\underline{\mbox{\footnotesize M}} = \ubar\mu, 
\underline{\mbox{\footnotesize N}}=\ubar\nu$, one gets instead 
generalized Gauss equations,
\be
({\cal R}_T)_{\ubar\mu\ubar \nu} = {\cal R}_{\ubar \mu\ubar\nu}
+ \Omega_{\ubar i \ubar\mu \ubar\rho} 
\Omega^{\ubar i}_{\;\;\, \ubar\nu \ubar\sigma} \theta^{\ubar\rho}
\wedge \theta^{\ubar\sigma} \;,
\label{Gauss}
\ee
relating the intrinsic curvature ${\cal R}_T$ on $M$ to the space-time
curvature ${\cal R}$. Analogously, taking 
$\underline{\mbox{\footnotesize M}} = \ubar i, 
\underline{\mbox{\footnotesize N}} =\ubar j$ in (\ref{CR}), 
gives generalized Ricci equations,
\be
({\cal R}_N)_{\ubar i\ubar j} = {\cal R}_{\ubar i\ubar j}
+ \Omega_{\ubar\mu \ubar\rho\ubar i}  
\Omega^{\ubar\mu}_{\;\;\,\ubar\sigma \ubar j} \theta^{\ubar\rho}
\wedge \theta^{\ubar\sigma} \;,
\label{Ricci}
\ee
relating the intrinsic curvature ${\cal R}_N$ on the normal 
bundle of $M$ to the space-time curvature ${\cal R}$.

The considerations discussed so far are general. 
We may now apply them to our particular situation, that is a trivial 
embedding ($\phi^i={\rm const.}$, $\phi^\mu=\sigma^\mu$) and backgrounds
satisfying the constraints (\ref{res2}). In this case, it can be shown 
(for instance going in normal coordinates) that the symmetric part of
the second fundamental form 
$\Omega^{\ubar i}_{\;\; (\ubar \mu \ubar\nu )}$ vanishes. Moreover, since 
$H_{\ubar\mu\ubar\nu\ubar\rho}=H_{\ubar\mu\ubar i\ubar j}=0$, one finds 
that $({\cal R}_T)_{\ubar\mu\ubar \nu} =(R_T)_{\ubar\mu\ubar \nu}=
R_{\ubar \mu\ubar\nu}$ and $({\cal R}_N)_{\ubar i\ubar j} =
(R_N)_{\ubar i\ubar j} = R_{\ubar i \ubar j}$. Therefore, defining the 
one-forms 
$H_{\ubar\mu \ubar i} = H_{\ubar\mu \ubar i \ubar \rho} \theta^{\ubar\rho}$, 
equations (\ref{Gauss}) and (\ref{Ricci}) yield the following simple 
expressions for the generalized curvatures of the tangent and the normal 
bundles:
\bea
\a\a {\cal R}_{\ubar \mu\ubar\nu} = R_{\ubar\mu\ubar\nu} -
H_{\ubar i \ubar\mu} \wedge H^{\ubar i}_{\;\;\,\ubar\nu} \;, \label{Gauss1} \\
\a\a {\cal R}_{\ubar i\ubar j} = R_{\ubar i\ubar j}
- H_{\ubar\mu \ubar i} \wedge H^{\ubar\mu}_{\;\;\,\ubar j} \;.
\label{Ricci1}
\eea

\end{document}